\documentclass[twocolumn,showpacs,floats,floatfix,superscriptaddress,aps,pra]{revtex4-1}

\usepackage{amsfonts}
\usepackage{amssymb}
\usepackage{amsmath}
\usepackage{calc}
\usepackage{graphicx}
\usepackage{bm}

\usepackage[normalem]{ulem}
\usepackage{amsmath,amssymb}
\usepackage{multirow}
\usepackage{xcolor,soul}
\usepackage{srcltx}
\usepackage{hyperref,graphicx}

\def\be{ \begin{equation}}
\def\ee{ \end{equation}}
\def\bea{ \begin{eqnarray}}
\def\eea{ \end{eqnarray}}
\def\bse{ \begin{subequations}}
\def\ese{ \end{subequations}}
\def\bc{ \begin{center}}
\def\ec{ \end{center}}

\begin{document}

\author{Stefano Longhi$^{*}$} 
\affiliation{Dipartimento di Fisica, Politecnico di Milano, Piazza L. da Vinci 32, I-20133 Milano, Italy}
\affiliation{IFISC (UIB-CSIC), Instituto de Fisica Interdisciplinar y Sistemas Complejos, E-07122 Palma de Mallorca, Spain}
\author{Ermanno Pinotti} 
\affiliation{Dipartimento di Fisica, Politecnico di Milano, Piazza L. da Vinci 32, I-20133 Milano, Italy}
\email{stefano.longhi@polimi.it}

\title{Non-Hermitian invisibility in tight-binding lattices}
  \normalsize


%
\bigskip
\begin{abstract}
\noindent  
A flexible control of wave scattering in complex media is of relevance in different areas of classical and quantum physics. Recently, a great interest has been devoted to scattering engineering in non-Hermitian systems, with the prediction and demonstration of  new classes of non-Hermitian potentials  with unique scattering properties,  such as transparent and invisibile potentials  or one-way reflectionless potentials. 
Such potentials have been found for both continuous and discrete (lattice) systems. However, wave scattering in lattice systems displays some distinct features arising from the discrete (rather than continuous) translational invariance of the system, characterized by a finite band of allowed energies and a finite speed of wave propagation on the lattice. Such distinct features can be exploited to realize invisibility on a lattice with methods that fail when applied to continuous systems. Here we show that a wide class of time-dependent non-Hermitian scattering potentials or defects with arbitrary spatial shape can be synthesized in an Hermitian single-band tight-binding lattice, which are fully invisible owing to the limited energy bandwidth of the lattice. 
 \end{abstract}

\maketitle

\section{Introduction}

In recent years there has been a surge of interest in
both classical and quantum systems that are described by effective non-Hermitian (NH)
Hamiltonians \cite{U1,U2,U3}. In such 
systems, wave transport, localization and scattering can
be deeply modified as compared to Hermitian systems.  In particular, a local NH scattering potential displaying spatial regions with gain and loss, which serve as sources and sinks
for waves, can be suitably engineered to control fundamental wave effects, such as interference and diffraction, in ways  that are impossible to
realize with conventional Hermitian systems. 
Suppressing wave scattering, thus
realizing transparency effects in inhomogeneous media,
is known since long time for Hermitian potentials \cite{r2,r6,r7,r9}.
However, most amazing effects, such as invisibility, could be realized only when considering NH potentials.
 Recently, wave reflection and scattering from complex potentials has sparked a great interest with the prediction of intriguing phenomena, such as unidirectional or bidirectional invisibility of the potential \cite{r9a,r9b,r10,r11,r12,r13,r13a,r14,r15,r16,r16c,r19b,r17,r17b}, asymmetric scattering \cite{r16b,noo,noo1}, constant-intensity wave transmission across suitably-engineered NH scattering landscapes \cite{A1,A2,A3,A4,A5,A6},
 reflectionless transmission based on the spatial Kramers-Kronig relations \cite{r18,r19,r20,r21,r22,r23,r24,r25,r26,r26bis,r27,r27bis,K1,K2,K3,K4}, and NH transparency \cite{K5}.\\
In continuous media, wave reflection is  usually described in terms of continuous wave equations both in space and time, such as  the Helmholtz equation or the stationary Schr\"odinger equation. However, in several physical systems, such as in quantum or classical transport on a lattice \cite{r28,r29,r30,r31,r32} or  in so-called discrete quantum mechanics \cite{r33,r34,r35}, space is discretized and wave transport is better described by the discrete version of the Schr\"odinger equation, where the the kinetic energy operator  $p_x^2$ is replaced by a periodic function $E(p_x)$ of the momentum operator $p_x$ that describes the dispersion band of the lattice.
Owing to the importance of such a broad class of discretized systems in different areas of physics, ranging from photonics to condensed-matter physics and beyond, scattering engineering in discrete models is becoming highly demanding  and could provide a fertile ground in many areas of science and
engineering in which discrete wave propagation is a key element.
Like for the continuous Schr\"odinger equation, reflectionless potentials can be constructed for the discrete Schr\"odinger equation as well \cite{r9a,r36,r38,r39,r40}, for example using the methods of supersymmetry for discrete systems \cite{r9a,r38,r39,r40}. Likewise, constant-intensity waves can be realized in suitably engineered complex lattices \cite{A4,A5,A6}, as demonstrated in a recent experiment \cite{A6}.
However, wave scattering in lattice systems display some distinct features, such as Bragg scattering, arising from the discrete (rather than continuous) translational invariance in space of the system, and characterized by a finite band of allowed energies and a finite speed of wave propagation on the lattice. Such distinct features can prevent the extension to discrete systems of wave scattering engineering methods valid for continuous systems. For example, it has been shown that the wide class of stationary Kramer-Kronig potentials, which are unidirectionally or bidirectionally reflectionless in continuous media, become reflective in discrete media owing to Bragg scattering \cite{K1}. On the other hand, the features of transport on a lattice arising from the discrete spatial invariance can be fruitfully exploited to realize forms of transparency that would be prevented in continuous systems. For example, since the speed of propagation on a lattice has an upper bound $v_m$ (according to the Lieb-Robinson bound \cite{Lieb}), any potential that drifts on the lattice at a speed $v$ faster than $v_m$ is necessarily reflectionless \cite{K2,K2b}.\\ 
In this work we suggest a simple method to synthesize space-time invisible potentials and defects of {\em arbitrary} spatial shape in tight-binding lattices that exploits the limited energy (frequency) bandwidth of the lattice, 
and propose a feasible photonic setup for the realization of such a broad class of invisible potentials. The method strictly works for a system with a bounded energy spectrum and thus it fails
when applied to continuous system, where energy is unbounded.  
 
\section{Wave scattering on a lattice}
\subsection{Model}
Let us consider wave scattering from a time-dependent NH potential or defect on a one-dimensional single-band tight-binding lattice, which in physical space is described by the coupled equations for the wave amplitudes $\psi_n(t)$ at various lattice sites $n$
\begin{equation}
i \frac{d\psi_n}{d t}=- \sum_l \kappa_{n-l}\psi_l+\sum_l V_{n,l}(t) \psi_l
\end{equation}
where $\kappa_l=\kappa_{-l}^*$ is the hopping amplitude between lattice sites distant $|l|$ on the lattice and the perturbation matrix $V_{n.l}(t)$ describes the time-dependent scattering potential or lattice defects, which is assumed to vanish fast enough as $|n,l| \rightarrow \infty$. For example, for a local on-site scattering potential the matrix $V_{n,l}(t)$ is diagonal, while for defects of the hopping amplitudes the matrix $V_{n,l}(t)$ contains off-diagonal non-vanishing elements. As for the time dependence of the perturbation matrix, special cases are those of a stationary (i.e. time-independent) or time-periodic potentials. However, we assume here a rather general dependence of time with the only constraint that $|V_{n,l}(t)|$ is a bounded function of time, i.e. it does not secularly grow in time, and can be expanded as a Fourier integral or generalized Fourier integral (i.e. containing undamped harmonic terms).

 In the absence of the scattering potential, $V_{n,l}(t)=0$, the eigenstates of Eq.(1) are extended Bloch waves, $\psi_n(t)=\exp[iqn-iE(q)t]$, with energy $E=E(q)$ defined by the dispersion relation
\begin{equation}
E(q)=- \sum_{l} \kappa_l \exp(-iql).
\end{equation}
A wave packet with carrier Bloch wave number $q$ propagates on the lattice with a group velocity $v_g=(dE/dq)$. For a lattice with short-range hopping, i.e. when $|\kappa_l| \rightarrow 0$ fast enough as $|l| \rightarrow \infty$, the group velocity displays un upper bound, according to Lieb and Robinson \cite{Lieb}. Likewise, the energy band displays a finite width $\Delta$. For example, for a tight-binding lattice with nearest-neighbor hopping amplitude, $\kappa_l=0$ for $l \neq \pm1$ and $\kappa_l= \kappa$ for $l= \pm 1$, the tight-binding dispersion curve reads $E(q)=-2 \kappa \cos q$, the energy band has a finite width given by $\Delta=4 \kappa$, and the group velocity $v_g=2 \kappa \sin q$ displays the upper bound $v_m=2 \kappa = \Delta /2$.\\

\subsection{Scattering analysis}
Let us now consider a spatially-localized time-dependent scattering potential and/or lattice defects, described by the time-dependent perturbation matrix $V_{n,m}(t)$ with $V_{n,m}(t) \rightarrow 0$ fast enough as $|n,m| \rightarrow \infty$. In particular, for a local on-site scattering potential $V_n(t)$ the perturbation matrix $V_{n,m}(t)$ is diagonal and given by $V_{n,m}(t)=V_n(t) \delta_{n,m}$. We assume that a Bloch (plane) wave with wave number $q$, energy $E=E(q)$ and positive group velocity $v_g>0$, coming from $n=-\infty$, is incident from the left side toward the scattering region. We can write the solution to Eq.(1) in the form
\begin{equation}
\psi_n(t)=\exp(iqn-iEt)+ \phi_n(t) \exp(-iEt)
\end{equation}
where the former term on the right hand side of Eq.(3) is the incoming plane wave while $\phi_n(t)$ are the amplitudes of scattered wave on the lattice, which satisfy the coupled equations
\begin{equation}
i \frac{d \phi_n}{dt}= -E \phi_n +\sum_l \left\{ V_{n,l}(t) - \kappa_{n-l}   \right\} \phi_l+ \sum_l V_{n,l}(t) \exp (iql).
\end{equation}
Equation (4) is a linear non-autonomous system with a forcing term in the variables $\phi_n(t)$ and should be solved with the appropriate boundary conditions, which depend rather generally on the time-dependence of the scattering perturbation matrix $V_{n,m}(t)$. For a static (time-independent) potential, the system is autonomous, $\phi_n(t)$ is independent of $t$ and outgoing boundary conditions should be imposed.  The same holds for a time-periodic potential, where Floquet analysis can be used \cite{r16c}. On the other hand, for arbitrary time-dependence of the potential Eq.(4) should be solved by considering the initial-value condition $\phi_n(-t_0)=0$ at some remote time $t_0 \rightarrow \infty$. Here we will consider this rather general case. In this case the NH scattering matrix $V_{n,m}(t)$ turns out to be invisible provided that, for any arbitrary incident wave, one has $\phi_n(t) \rightarrow 0$ for the scattered wave amplitudes as $|n| \rightarrow \infty$ for any time instant $t$.

\subsection{The continuous (long-wavelength) limit of wave scattering}
In the limit of a tight-binding lattice with nearest-neighbor hopping of amplitude $\kappa$ and for an on-site scattering potential $V_n(t)$, i.e. $V_{n,m}(t)=V_n(t) \delta_{n,m}$, we can solve Eq.(1) by letting $\psi_n(n)=\psi(x=n,t)$, where the wave function $\psi(x,t)$ of continuous space and time variables $x$ and $t$ satisfies the discrete Schr\"odinger equation (see for instance \cite{K1,r41})
\begin{equation}
i \frac{\partial \psi}{\partial t} =- 2 \kappa \cos (p_x) \psi(x,t)+ V(x,t) \psi(x,t)
\end{equation}
where $p_x=-i \partial_x$ and $V_n(t)=V(x=n,t)$. \\
The kinetic energy term in Eq.(5) is periodic in the momentum $p_x$, which implies that a limited interval of energies are allowed in a lattice system, corresponding to propagative (Bloch) waves. The invisibility method that will be presented in the next section works provided that the kinetic energy operator is bounded, like in a system with discrete translational invariance, but breaks down when the kinetic energy term is unbounded from above, like in a system with continuous translational symmetry. This case is found in the so-called long-wavelength limit of the discrete Schr\"odinger equation, which corresponds to low-energy excitation of the system. Specifically,
for a potential $V(x,t)$ that varies slowly with respect to $x$ over the lattice period and for low-energy excitation of the system, we may expand the kinetic energy operator $\cos (p_x)$ in the neighbor of $p_x=0$, i.e. we may let $\cos (p_x) \simeq 1-p_x^2/2$ (long-wavelength approximation). In this case, omitting an inessential constant energy potential term, Eq.(5) reads
\begin{equation}
i \frac{\partial \psi}{\partial t} =-  \kappa \frac{\partial^2 \psi}{\partial x^2}+ V(x,t) \psi(x,t)
\end{equation}
which is the continuous Schr\"odinger equation describing the scattering of a quantum particle in the NH time-dependent potential $V(x,t)$. In this limit we loose the discrete translational invariance of the lattice and the energy of the system becomes unbounded from above. 

\section{Non-Hermitian invisibility}
\subsection{The general result}
NH invisibility in lattice systems has been predicted in some previous works, including nearest-neighbor tight-binding lattices under special stationary potentials synthesized by the methods of supersymmetry \cite{r9a} or
for harmonically-oscillating on-site potentials  \cite{r16c}, and for the class of Kramers-Kronig potentials drifting on the lattice at a speed faster than $v_g$ \cite{K2}. The last method exploits the finite speed of wave propagation in the lattice, so that any potential that drifts on the lattice at a speed faster than $v_m$ cannot reflect waves and thus is transparent. 

Here we widen the class of NH invisible potentials on a lattice, beyond the nearest-neighbor approximation and including also scattering from off-diagonal (hopping) defects in addition to on-site (diagonal) potential, exploiting the finiteness of the energy bandwidth of the lattice. The main physical idea is as follows. Let us assume that the Fourier spectrum $\hat{V}_{n,m}(\omega)$ of the perturbation scattering matrix $V_{n,m}(t)$, defined by
\begin{equation}
\hat{V}_{n,m}(\omega)=\int_{-\infty}^{\infty} dt V_{n,m}(t) \exp(i \omega t)
\end{equation}
 vanishes for all frequencies $\omega< \omega_0$ (or, likewise, for any frequency $\omega> -\omega_0$), where $\omega_0$ is larger than the bandwidth $\Delta $ of the tight-binding lattice. The interaction of the incidente wave with the time-varying potential is inelastic and involves the absorption or emission of energy quanta from the oscillating potential \cite{uffa1,uffa2}. However, since the Fourier spectrum of the oscillating potential is composed by $\omega>\omega_0$ frequency components solely, the energies $E^{'}$ of the scattered wave are constrained by the inequality  \cite{r16c,uffa3} $E^{'}=E+\omega>E+\omega_0>E+\Delta$, i.e. they fall outside the allowed energy band of the lattice. Therefore, far from the scattering region where the lattice is uniform, the scattered waves are Bloch waves but with an imaginary Bloch wave number, i.e. they are evanescent waves decaying toward zero as $|n| \rightarrow \infty$. This means that the scattered waves cannot be propagative in the homogeneous lattice regions, far from the scattering potential, which clearly implies invisibility. \\
 We emphasize that this result holds regardless of the specific spatial shape of the potential and specific shape of incoming waves, so that the kind of invisibility induced by the temporal modulation does not require any special tailoring in space of the scattering potential nor special initial excitation of the system with prescribed input state (such as e.g. in Ref.\cite{A6}).
From this simple physical picture, we can conclude that the following general property holds:
 \\{\em Any scattering NH matrix perturbation $V_{n,m}(t)$ such that its Fourier spectrum vanishes for frequencies $\omega<\omega_0$ (or likewise for frequencies with $\omega>-\omega_0$), with $\omega_0$ larger than the width $\Delta$ of the tight-binding lattice band, is invisible}\\
 To prove the above general property, let us integrate the coupled equations (4) for the scattered wave amplitudes $\phi_n(t)$ at various lattice sites with the initial condition $\phi_n(-t_0)=0$ at a far remote time $t_0 \rightarrow \infty$ using a modified Laplace-Fourier method. 
 To this aim, let $f(t)$ be a regular function of time $t$, defined for $t \geq -t_0$ and bounded (or growing in time lower than any exponential) as $t \rightarrow \infty$. For a given time $t_1>0$, arbitrarily large, let us introduce the modified Fourier-Laplace spectrum
 \begin{equation}
 \hat{f}^{(\epsilon)}(\omega)=\int_{-t_0}^{t_1} dt f(t) \exp(i \omega t-\epsilon t)
 \end{equation}
 where $\epsilon>0$ is a small positive number. Note that the previous relation reduces to the usual Fourier spectrum in the limit $\epsilon=0$ and $t_1,t_0 \rightarrow \infty$. The inverse relation to Eq.(8) reads (see Appendix)
 \begin{equation}
 f(t)=\frac{1}{2 \pi }\exp( \epsilon t)  \int_{-\infty}^{\infty} d \omega \hat{f}^{(\epsilon)}(\omega) \exp(-i \omega t)
 \end{equation}
 which is valid for $-t_0<t<-t_1$.\\
 Here we are interested in considering the triple limit $t_0,t_1 \rightarrow \infty$, $\epsilon \rightarrow 0^+$ with $\epsilon t_0 \rightarrow 0$ and $\epsilon t_1 \rightarrow \infty$. Multiplying both sides of Eq.(4) by $\exp(i \omega t-\epsilon t)$ and integrating over time $t$ from $-t_0$ to $t_1$, taking into account that $\phi_n(-t_0) \exp(\epsilon t_0) \simeq \phi_n(-t_0)=0$ and $\phi_n(t_1) \exp(-\epsilon t_1) \simeq 0$ in the above mentioned limit, one obtains
 \begin{eqnarray}
 (E+\omega+i \epsilon) \hat{\phi}_{n}^{(\epsilon)} (\omega)+\sum_l \kappa_{n-l}\hat{\phi}_{l}^{(\epsilon)} (\omega) & + \nonumber \\
 - \frac{1}{2 \pi} \sum_l \int_{-\infty}^{\infty} d \Omega \hat{V}_{n,l}(\Omega) \hat{\phi}_{l}^{(\epsilon)}(\omega-\Omega) & = \\
 \sum_l \hat{V}_{n,l}^{(\epsilon)} (\omega) \exp (iql). \nonumber
 \end{eqnarray}
 In deriving Eq.(10), we used the relation 
 \begin{eqnarray}
 \int_{-t_0}^{t_1} dt V_{n,l}(t) \phi_l(t) \exp(i \omega t -\epsilon t) \nonumber \\
 \simeq \frac{1}{2 \pi}  \int_{-\infty}^{\infty} d \Omega   \hat{V}_{n,l}(\Omega)  \hat{\phi}_l^{(\epsilon)} (\omega-\Omega)
 \end{eqnarray}
 which is valid in the large $t_0, t_1$ limit, as shown in the Appendix. Since $\hat{V}_{n,l}(\Omega)=0$ for $\Omega < \omega_0$, from Eq.(10) it readily follows that the spectral amplitude $\hat{\phi}_{n}^{(\epsilon)} (\omega) $ depends on all other spectral amplitudes  $\hat{\phi}_{l}^{(\epsilon)} (\omega^{\prime})$ at frequencies $\omega^{\prime}< \omega-\omega_0$. Moreover, in the large $t_0,t_1$ and small $\epsilon$ limits, $\hat{V}_{n,l}^{(\epsilon)} (\omega) \simeq \hat{V}_{n,l}(\omega)=0$ for $\omega>\omega_0$ (see Appendix), so that the 
 forcing term of the spectral amplitude  $\hat{\phi}_{n}^{(\epsilon)} (\omega) $ [the term on the right hand side of Eq.(10)] vanishes for $\omega< \omega_0$. This implies that 
\begin{equation}
\hat{\phi}_{n}^{(\epsilon)} (\omega) =0
\end{equation}
 for any frequency $\omega \leq \omega_0$, in agreement with the physical picture that the scattered waves cannot transport energies (frequencies) smaller than $E+\omega_0$. Therefore we can write
  \begin{equation}
 \phi_n(t)=\frac{1}{2 \pi }\exp( \epsilon t)  \int_{\omega_0}^{\infty} d \omega \hat{\phi}_{n}^{(\epsilon)}(\omega) \exp(-i \omega t).
 \end{equation}
Let us now consider the behavior of $\phi_n(t)$ as $n \rightarrow \pm \infty$, i.e. far from the scattering region. In this limit, the spectral amplitudes $\hat{\phi}_{n}^{(\epsilon)}(\omega) $ with $\omega> \omega_0$  satisfy the linear dispersion equation
\begin{equation}
 (E+\omega+i \epsilon) \hat{\phi}_{n}^{(\epsilon)} (\omega)+\sum_l \kappa_{n-l}\hat{\phi}_{l}^{(\epsilon)} (\omega)=0
 \end{equation}
which is obtained from Eq.(10) by neglecting the vanishing scattering matrix elements as $n \rightarrow \pm \infty$. The solution to Eq.(14) is given in terms of superposition of Bloch waves, namely $\hat{\phi}_{n}^{(\epsilon)} (\omega) \sim Y_{\pm}(\omega) \exp(iQ_{\pm}n)$ as $n \rightarrow \pm \infty$ with some complex amplitudes $Y_{\pm}(\omega)$ and {\em complex} Bloch wave numbers $Q_{\pm}=Q_{\pm}(\omega)$. The complex Bloch wave numbers are obtained as a solution of the dispersion equation
\begin{equation}
E+\omega=E(Q_{\pm})=-\sum_n \kappa_n \exp(-iQ_{\pm}n).
\end{equation}
with ${\rm Im}(Q_+)<0$ and ${\rm Im}(Q_-)>0$
Since the energy $E+\omega>E+\Delta$ falls outside the band, the imaginary parts of $Q_{\pm}$ are strictly nonvanishing and thus the Bloch waves $\hat{\phi}_{n}^{(\epsilon)} (\omega)$ are {\em evanescent}, exponentially decaying as $n \rightarrow  \pm \infty$ for any frequency $\omega>\omega_0$. Indicating by $Q_{+,m}>0$ the minimum of $| {\rm Im}(Q_+)|$ over the range of frequencies $\omega \geq \omega_0$, from Eq.(13) for $ n \rightarrow \infty$ one has
\begin{eqnarray}
|\phi_n(t)|  \sim \frac{1}{2 \pi} \exp( \epsilon t)  \left| \int d \omega Y_+(\omega) \exp(i \omega t-i Q_+ n) | \right| \nonumber \\
 < \frac{1}{2 \pi} \exp( \epsilon t)  \left| \int d \omega Y_+(\omega) \exp(i \omega t)  \right| \exp(-Q_{+,m } n) \;\;\;\;\;
\end{eqnarray}
and thus $|\phi_n(t)| \rightarrow 0$ as $n \rightarrow \infty$ for any time instant $t$. Likewise, one has $|\phi_n(t)| \rightarrow 0$ as $n \rightarrow -\infty$ for any time instant $t$, indicating that the NH scattering perturbation matrix is invisible.

Clearly, the invisibility property strictly requires a finite bandwidth of allowed energies in the system, and thus it breaks down in the continuous (long-wavelength) limit of lattice dynamics. In fact, in this limit the dynamics can be described by a usual
continuous Schr\"odinger equation [Eq.(6)] with an unbounded range of allowed energies. Since the energies $E^{\prime}$ of scattered waves can now fall into allowed energy intervals,  they are now propagative (rather than evanescent) waves, and the invisibility property is thus lost: Only for some special tailored space-time potentials scattering can be prevented (see for instance \cite{r26bis}).

\begin{figure}[h]
   \centering
    \includegraphics[width=0.49\textwidth]{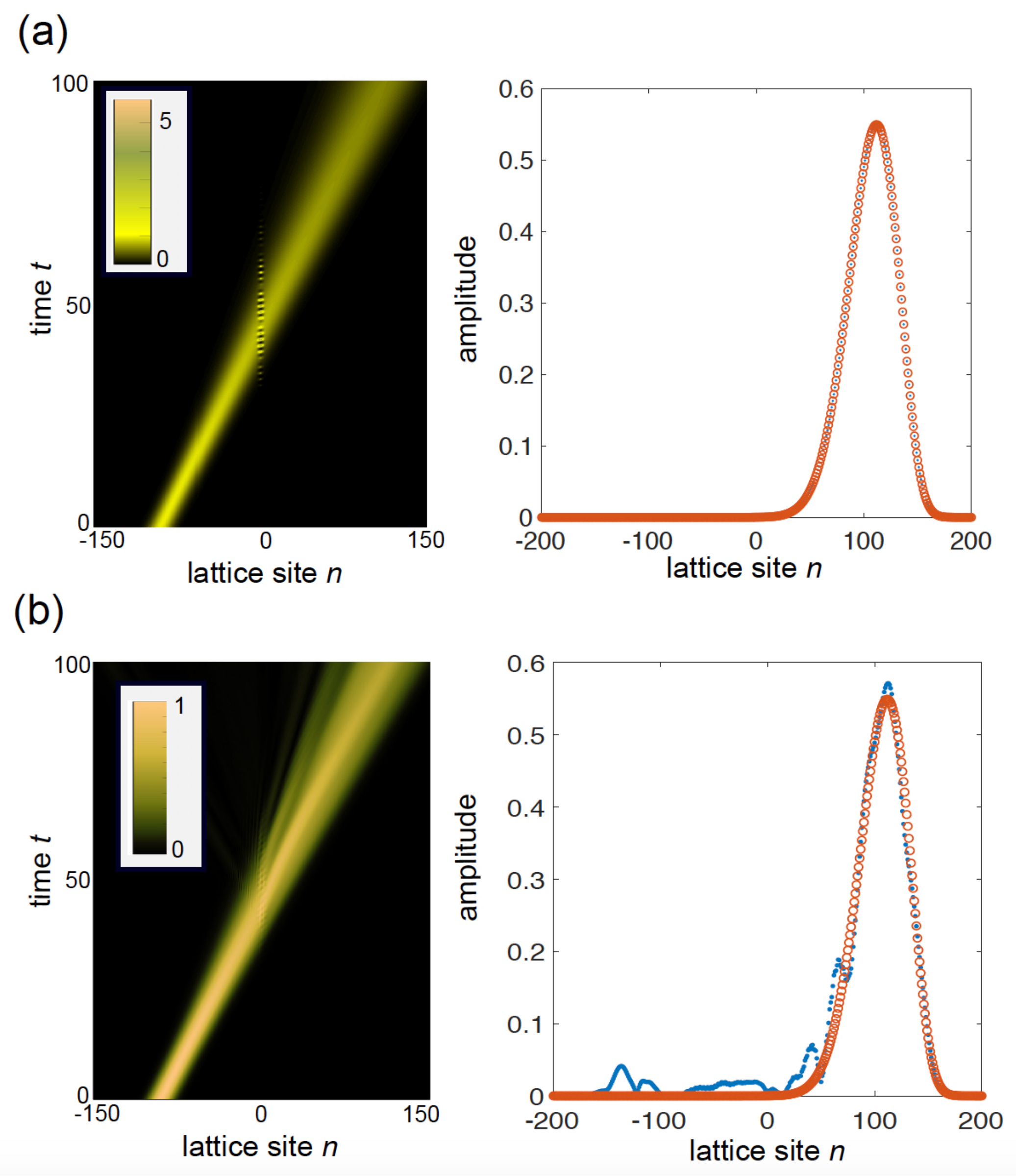}
    \caption{Numerically-computed propagation of a Gaussian wave packet in a tight-binding lattice with nearest and next-to-nearest neighbor hopping across a time-varying local scattering potential $V_{n,m}(t)=V_n R(t) \delta_{n,m}$. The scattering potential is $V_n=V_0 \exp(-x^2/w^2)$;  the modulation function is $R(t)=A_1 \exp(i \omega_1 t)+A_2 \exp(i \omega_2 t)$ in (a), and $R(t)=A_1 \cos ( \omega_1t)+ A_2 \cos (\omega_2 t)$ in (b), with incommensurate frequencies $\omega_1$ and $\omega_2$. Parameter values are given in the text. The lattice is excited at initial time $t=0$ with the Gaussian wave distribution $\psi_n(0)=\exp \{ -[(n+90)/10]^2 +i q n \}$ with carrier Bloch wave number $q= \pi/2$. The left panels show on a pseudo-color map the time evolution of the amplitudes $|\psi_n(t)|$, whereas the right panels show the behavior of the amplitudes $|\psi_n(t_0)|$ at final time $t_0=100$ (small blue dots), after the wave packet has fully crossed the scattering region. The open red circles in panels (b) depict the behavior of $|\psi_n(t_0)|$ that one would observe in the absence of the scattering potential, i.e. for the freely-moving wave packet. Note that in (a) the two curves are overlapped, indicating that the scattering potential is invisible.}
\end{figure}

\begin{figure}[h]
   \centering
    \includegraphics[width=0.49\textwidth]{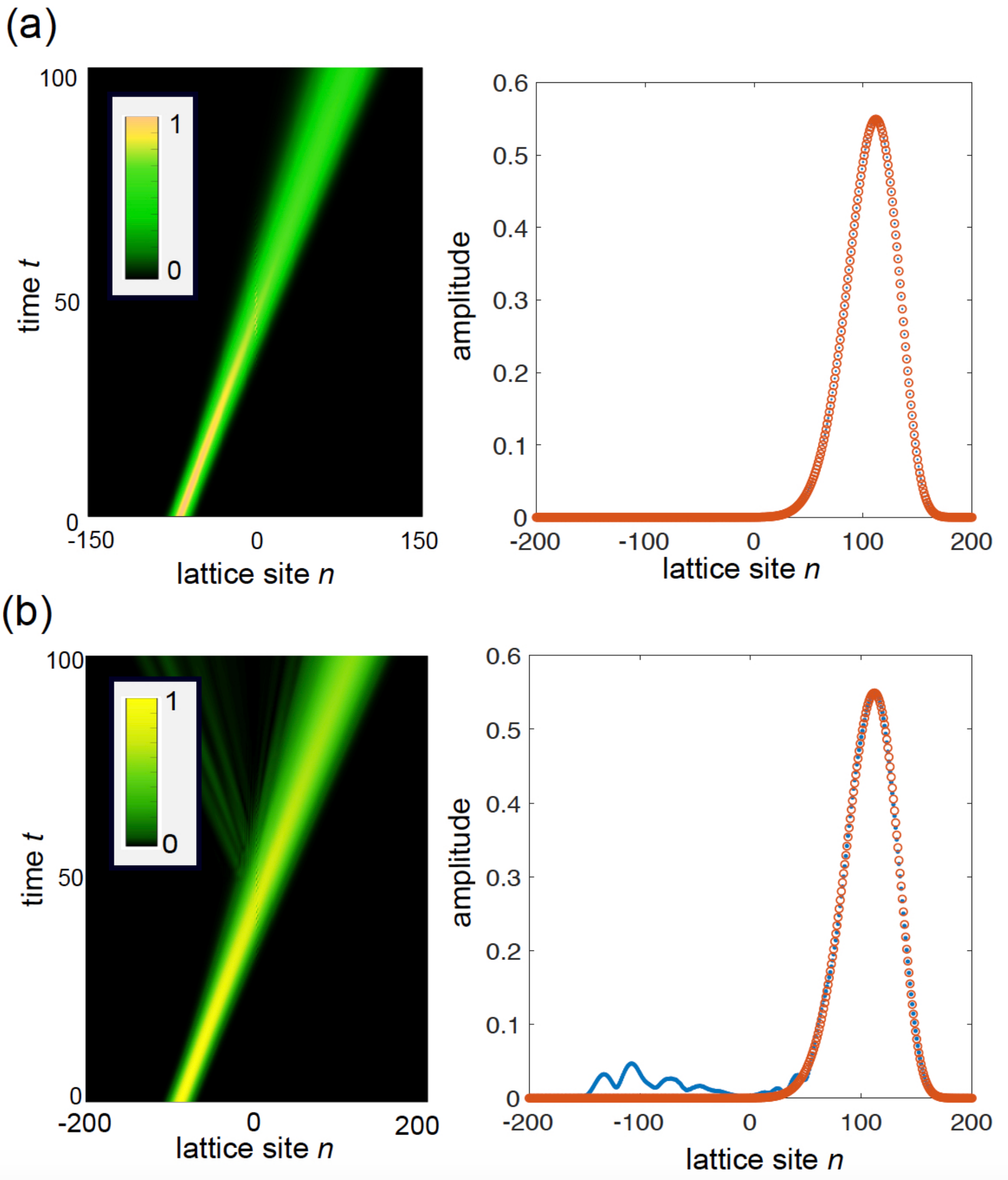}
    \caption{Same as Fig.1, but for a perturbation scattering matrix describing a time-varying hopping defect $V_{n,m}=R(t) (\delta_{n,0}\delta_{m,1}+\delta_{n,1} \delta_{m,0})$ between sites $n=0$ and $n=1$. Parameter values are given in the text.}
\end{figure}

\subsection{Illustrative examples}
To exemplify the main result presented in the previous subsection and to check the correctness of the theoretical analysis, let us consider a scattering matrix which can be factorized as
\begin{equation}
V_{n,m}(t)=R(t) T_{n,m}
\end{equation}
with a time-independent matrix $T_{n,m}$ and a function of time $R(t)$. An invisible potential is obtained, for example, by  assuming for $R(t)$ a superposition of positive-frequency harmonics, i.e. $R(t) = \sum_{\alpha} A_{\alpha} \exp(i \omega_{\alpha} t)$, with frequencies $\omega_{\alpha} > \Delta$. The frequencies $\omega_{\alpha}$ could be rather generally incommensurate, so as a standard Floquet analysis of inelastic scattering \cite{r16c,uffa1} cannot be applied. Yet our general analysis discussed in previous subsection predicts invisibility of the scattering potential. As for the form of the matrix $T_{n,m}$, we consider two typical cases. The first one corresponds to a local on-site scattering potential, i.e. $T_{n,m}=V_n \delta_{n,m}$, while the second case corresponds to a defect in the hopping amplitudes of the lattice, for example we can assume $T_{n,m}=\delta_{n,0} \delta_{m,1}+\delta_{n,1} \delta_{m,0}$, which corresponds to modify the hopping amplitude between sites $n=0$ and $n=1$ of the lattice from $\kappa_1$ to $\kappa_1+R(t)$.

Figures 1 and 2 demonstrate the invisibility of the oscillating scattering perturbation in the two cases. The figures depict the evolution of a forward-propagating Gaussian wave packet, which is scattered off by the time-varying potential. At initial time $t=0$ the wave packet is localized at the left side far from the scattering region, and propagates forward toward the scattering region.  Coupled equations (1) have been numerically solved using an accurate variable-step fourth-order Runge-Kutta method. We assumed a tight-binding lattice with nearest and next-to-nearest neighbor hopping amplitudes $\kappa_1=\kappa_{-1}=1$, $\kappa_2=\kappa_{-2}=0.2$ and $\kappa_l=0$ for $l \neq  \pm 1, \pm 2$. The bandwidth of the lattice is $\Delta=4$. In Fig.1 we have a local scattering potential $V_n=V_0 \exp(-n^2/w^2)$ of amplitude $V_0=5$ and size $w=2$, while in Fig.2 we have a defect of the hopping amplitude between sites $n=0$ and $n=1$, described by the perturbation matrix $T_{n,m}=\delta_{n,0} \delta_{m,1}+\delta_{n,1} \delta_{m,0}$. Two different modulation functions $R(t)$ have been used, namely $R(t)=A_1 \exp(i \omega_1 t)+A_2 \exp(i \omega_2 t)$ [Figs.1(a) and 2(a)], and  $R(t)=A_1 \cos( \omega_1 t)+A_2 \cos( \omega_2 t)$ [Figs.1(b) and 2(b)]. Parameter values are $A_1=A_2=1$, $\omega_1=5$, and $\omega_2= \sqrt{18}$. Note that in the former case the modulation function has a positive-frequency spectrum and, since $\omega_{1,2}>\Delta$, invisibility  is clearly observed according to the theoretical analysis. Conversely, in the latter case the Fourier spectrum of $R(t)$ is bilateral, corresponding to an Hermitian perturbation, and invisibility is not anymore observed. 
\subsection{Physical implementation}
Synthetic lattices based on photonic, mechanical, acoustic, electrical or ultracold atomic systems could provide possible physical platforms for the observation of the invisibility effect predicted in this work. Here we discuss in details a possible experimental setup based on  photonic quantum walks of light pulses in coupled fiber loops \cite{R28}. Photonic quantum walks realize a synthetic lattice in time domain, enabling a flexible control of non-Hermitian terms in the Hamiltonian. They have provided recently a fascinating platform to experimentally access a wealth of novel non-Hermitian phenomena,  including parity-time symmetry breaking \cite{R29,R30}, non-Hermitian topological physics \cite{R31,R32,R33,R34,R34b}, non-Hermitian Anderson localization \cite{R35}, constant-intensity waves and induced transparency in complex scattering potential \cite{A6}, and multiple non-Hermitian phase transitions \cite{R36}. The system consists of two fiber loops of slightly different lengths  $L \pm \Delta L$ (short and long paths) that are connected by a fiber coupler with a coupling angle $\beta$. Two synchronized amplitude and phase modulators are placed in one of the two loops to control on demand the amplitude and phase of the traveling pulses at each transit. The traveling times of light in the two loops are $T \pm \Delta T$, where $T=L/c$,  $c$ is the group velocity of light in the fiber at the probing wavelength, and $\Delta T= \Delta L/c \ll T$ is the time mismatch arising from fiber length unbalance. The light dynamics of the optical pulses at successive transits in the two loops is considered at discretized times $t=t_n^m=n \Delta T+m T$, where $n=0, \pm 1, \pm2 ,...$ defines the site number of the synthetic lattice at various time slots and $m$ is the round-trip number, assumed to match the traveling time $T$ along the mean path length $L$.
Indicating by $u_n^{(m)}$ and $v_n^{(m)}$ the field amplitudes at the discretized times $t_n^m$ of the pulses in the two loops, 
 light dynamics in the coupled fiber loops is governed by the discrete-time coupled equations (see e.g. \cite{
A6,K2b,R30,R33,R34b,R36,R37})
 \begin{eqnarray}
 u^{(m+1)}_n & = & \left[   \cos \beta u^{(m)}_{n+1}+i \sin \beta v^{(m)}_{n+1}  \right]  \exp [-2iV_n^{(m)}] \;\;\;\;
 \\
 v^{(m+1)}_n & = & \left[   \cos \beta v^{(m)}_{n-1}+i \sin \beta u^{(m)}_{n-1}  \right]
 \end{eqnarray}
  where $V_n^{(m)}$ is the complex NH scattering potential that is realized by suitable control of synchronized phase and amplitude modulators. In the absence of the scattering potential, i.e. for $V_n^{(m)}=0$, the synthetic lattice  shows discrete spatial invariance and the corresponding Bloch eigenfunctions to Eqs.(18) and (19) are given by $(u_n^{(m)}, v_n^{(m)})^T=(\bar{u},\bar{v}) \exp[iqn-i E(q) m]$, where $q$ is the spatial Bloch wave number and $E(q)$ is the quasi energy. Owing to the binary nature of the lattice, two quasi-energy bands are found with dispersion relations $E_{\pm}(q)$ given by
 \begin{equation}
 E_{\pm}(q)= \pm {\rm acos} \left(  \cos \beta \cos q \right).
 \end{equation}
 \begin{figure}[h]
   \centering
    \includegraphics[width=0.49\textwidth]{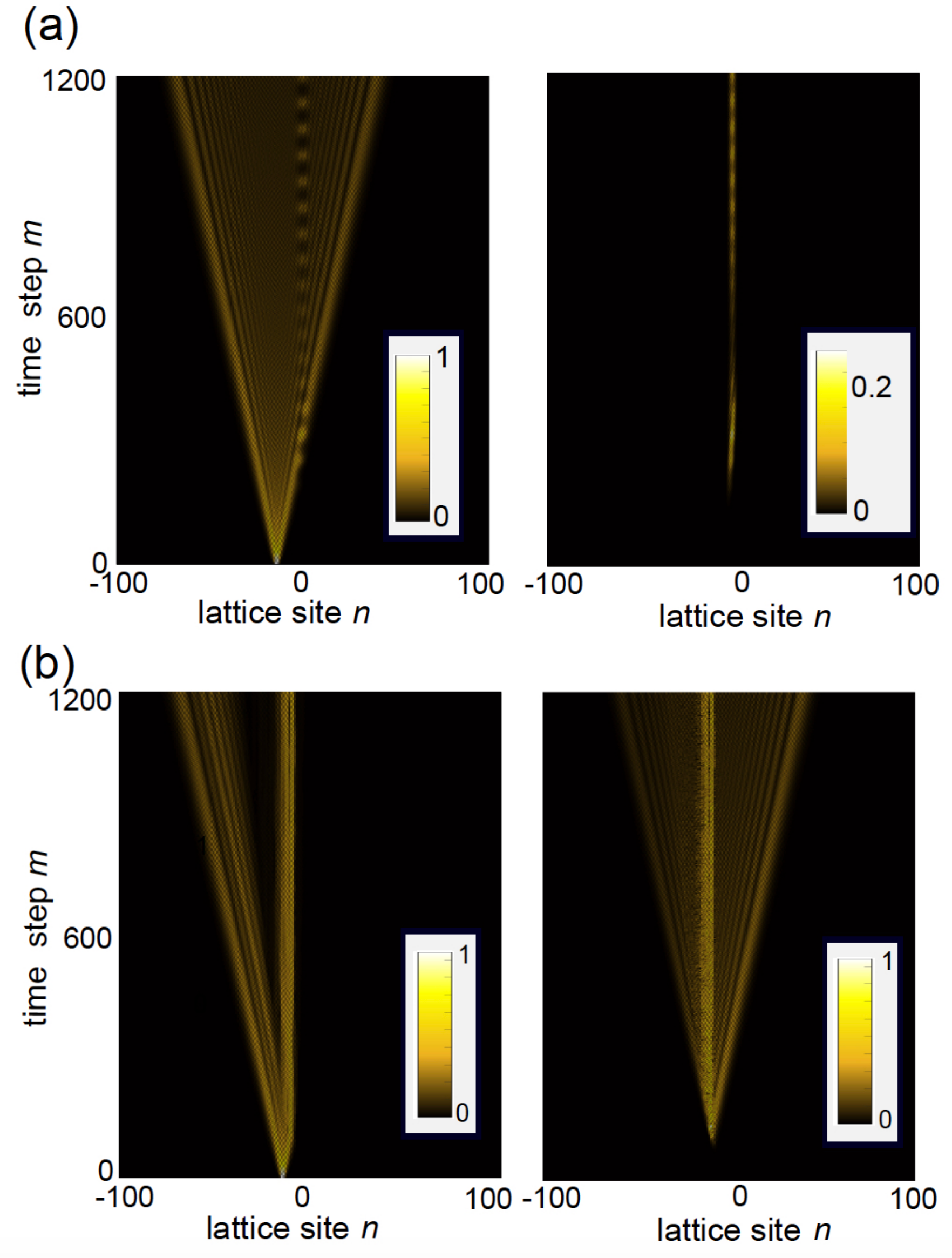}
    \caption{Invisible NH oscillating potentials in a photonic quantum walk. The scattering potential is $V_n^{(m)}=R_m \exp[ -(n/3)^2]$ with modulation function $R_m=A_1 \exp(i \omega_1 m)+A_2 \exp(i \omega_2 m)$ in (a), and  $R_m=A_1 \cos(\omega_1 m)+A_2 \cos( \omega_2 m)$ in (b). Parameter values are  $\beta= 0.97 \times \pi/2$ (coupling angle), $A_1=0.1$, $A_2=0.06$, $\omega_1=0.1$, and  $\omega_2= \sqrt{2} / 15 \simeq 0.0943$. Initial excitation condition of the system is $u_n^{(0)}=\delta _{n,-15}$ and $v^{(0)}_n=0$, corresponding to a single optical pulse injected into one of the two fiber loops at time slot (lattice site) $n=-15$.
   The left panels show the discrete-time evolution of the light intensity $(|u_n^{(m)}|^2+|v_n^{(m)}|^2)$ at various lattice sites $n$ on a pseudo color map, whereas the right panels show the evolution of 
   $I_n^{(m)} \equiv (|u_n^{(m)}-\tilde{u}_n^{(m)}|^2+|v_n^{(m)}-\tilde{v}_n^{(m)}|^2)$, where $\tilde{u}_n^{(m)}$ and $\tilde{v}_n^{(m)}$ are the pulse amplitudes in the fiber loops that one would observe in the absence of the scattering potential. In (a) one clearly sees that far from the scattering region one has $I_n^{(m) }\simeq 0$, indicating that the scattering potential in invisible. Conversely, in (b) the potential is not invisible.}
\end{figure}
Note that for a coupling angle $\beta$ close to $\pi/2$, i.e. for $\beta= \pi/2-\rho$ with $| \rho | \ll 1$, the dispersion relations of the two quasi energies read 
\begin{equation}
E_{\pm}(q)= \pm \frac{\pi}{2} \mp \rho \cos(q),
\end{equation}
i.e. they correspond to the shifted dispersion curves of two tight-binding lattices with nearest-neighbor hopping amplitude $\kappa=\pm \rho /2$.
In order to observe the NH invisibility predicted in this work, we consider  the continuous-time limit of the discrete-time quantum walk \cite{R34b}, which is obtained by assuming a coupling angle $\beta$ close to $\pi /2$, i.e.  $\beta=\pi/2-\rho$ with $| \rho | \ll 1$, and a small and slowly-varying amplitude of the scattering potential amplitude, i.e. $|V_n^{(m)}| \sim O(\rho) \ll 1$ and $V_n^{(m+1)} \simeq V_n^{(m)}$. 
At first order in $\rho$, Eqs.(18) and (19) take the form
  \begin{eqnarray}
 u^{(m+1)}_n & = & \left[   \rho u^{(m)}_{n+1}+i  v^{(m)}_{n+1}  \right]  \exp (-2iV_n^{(m)}) \\
 v^{(m+1)}_n & = & \left[   \rho v^{(m)}_{n-1}+i u^{(m)}_{n-1}  \right].
 \end{eqnarray}
 From the above equations, one can eliminate from the dynamics the variables $v_n^{(m)}$, yielding a second-order difference equation for $u_n^{(m)}$, which is solved by letting \cite{R34b}
 \begin{equation}
 u_{n}^{(m)}=(i)^m \left\{ \psi^{(+)}_n(m)+ (-1)^m \psi^{(-)}_n(m) \right\}.
 \end{equation}
 In Eq.(24), $\psi^{(\pm)}_n(m)$ are slowly-varying functions of the discrete time $m$ which satisfy the decoupled continuous-time Schr\"odinger equations
 \begin{equation}
 i \frac{d \psi_n^{(\pm)}}{dt}=  \pm \kappa  \left( \psi_{n+1}^{(\pm)}+ \kappa  \psi_{n-1}^{(\pm)} \right) +V_n(t) \psi_n^{(\pm)}
 \end{equation}
where we have set $\kappa \equiv (\rho /2)$, $t=m$ (considered as a continuous variable) and $V_n(t)=V_n^{(m)}$. Therefore, in the continuous-time limit the light pulse dynamics in the photonic quantum walk setup emulates the scattering dynamics from a NH time-dependent potential $V_n(t)$ on two independent tight-binding lattices with nearest-neighbor hooping amplitudes $\pm \kappa$. Provided that the Fourier spectrum of the potential $V_n(t)$ vanishes for all frequencies $\omega$ smaller than the bandwidth $4 \kappa$ of the tight-binding lattices, the scattering potential turns out to be invisible. An illustrative example, showing an invisible potential in the quantum walk system, is shown in Fig.3. The figure depicts the numerically-computed light pulse dynamics in the coupled fiber loops, as obtained by solving the discrete-time coupled equations (18) and (19), i.e. without any approximation, for a coupling angle $\beta=0.97 \times \pi/2$ and for a scattering potential $V_n^{(m)}=R_m \exp[ -(n/3)^2]$ with modulation function $R_m=A_1 \exp(i \omega_1 m)+A_2 \exp(i \omega_2 m)$ in Fig.3(a), and  $R_m=A_1 \cos(\omega_1 m)+A_2 \cos( \omega_2 m)$ in Fig.3(b) ($A_1=0.1$, $A_2=0.06$, $\omega_1=0.1$, $\omega_2= \sqrt{2} / 15 \simeq 0.0943$). The system is initially excited, at time $m=0$, with a single pulse injected into one of the two loops at the site $n=-15$, far from the scattering potential, i.e. we assumed as an initial condition $u_n^{(0)}=\delta_{n,-15}$ and $v_n^{(0)}=0$. The initial excitation spreads along the lattice and is scattered off by the oscillating potential near the $n=0$ region.
In the former case, where the complex modulation function $R_m$ is composed by positive-frequency components solely, with frequencies larger than the width of the lattice band, the potential turns out to be invisible [Fig.3(a)], while in the latter case, corresponding to a real modulation function with positive and negative frequency components, the potential in not invisible and scattered propagative waves are clearly visible [Fig.3(b)].

\begin{figure*}
   \centering
    \includegraphics[width=0.95\textwidth]{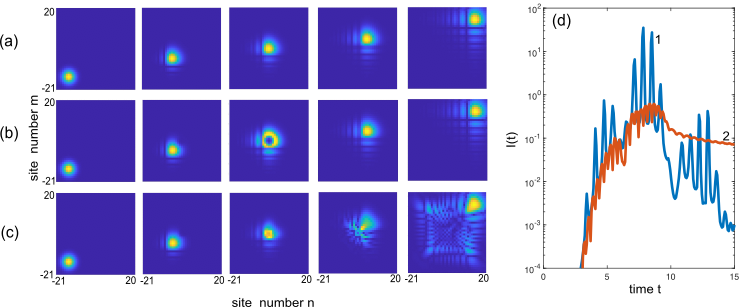}
    \caption{Scattering dynamics of a wave packet in a 2D square lattice from a space-time Gaussian-shaped on-site scattering potential. Parameter values are given in the text. The five panels in (a), (b) and (c) show on a pseudo-color map the numerically-computed evolution of the wave packet amplitudes $|\psi_{n,m}(t)|$ at a few increasing times $t$ ($t=0$, $t=5$, $t=7.5$, $t=10$ and $t=15$ from left to right).  In (a) there is not any scattering potential, in (b) there is the NH (invisible) space-time scattering potential, in (c) there is the Hermitian space-time scattering potential. (d) Temporal behavior of the error function $I(t)$, on a log scale, measuring the wave packet reconstruction of the wave packet after the scattering event. Curves 1 and 2 refer to the scattering by the non-Hermitian  and by the Hermitian potentials, respectively.}
\end{figure*}

\subsection{Invisibility in two-dimensional lattices}
The previous analysis has been focused to invisibility in one-dimensional lattice systems, however the results can be readily extended to scattering by local space-time perturbations in two-dimensional (2D) single-band lattices, namely any scattering NH perturbation such that its Fourier spectrum vanishes for frequencies $\omega<\omega_0$ (or likewise for frequencies with $\omega>-\omega_0$), with $\omega_0$ larger than the width $\Delta$ of the tight-binding lattice band, is invisible. In fact, the main physics underlying the invisibility property of such scattering potentials is that  the energies of the scattered waves fall outside the allowed energy band of the lattice. Therefore, far from the scattering region where the lattice is uniform, the scattered waves are Bloch waves but with an imaginary Bloch wave number, i.e. they are evanescent waves decaying toward zero at infinity. This result holds regardless of the spatial dimensionality of the lattice, indicating that invisibility is  observed also for 2D lattice systems. As an illustrative example, let us consider the scattering in a 2D square lattice with nearest-neighbor hopping amplitude $\kappa$ from a local NH on-site potential $V_{n,m}(t)$, which is described by the discrete Schr\"odinger equation
\begin{eqnarray}
i \frac{d\psi_{n,m}}{d t} & = & - \kappa \left( \psi_{n+1,m}+\psi_{n-1,m}+\psi_{n,m+1}+\psi_{n,m-1}  \right)  \nonumber \\
& + & V_{n,m}(t) \psi_{n,m}
\end{eqnarray}
where $\psi_{n,m}(t)$ is the wave amplitude at site $(n,m)$ of the square lattice.  As in Sec.III.B
we consider a scattering potential of the form. $V_{n,m}(t)=R(t) V_{n,m}$ 
with a time-independent matrix $V_{n,m}$, defining the spatial shape of the local 2D scattering potential, and a modulation function $R(t)$ given by either
$R(t)=A_1 \exp(i \omega_1 t)+A_2 \exp(i \omega_2 t)$ or $R(t)=A_1 \cos ( \omega_1t)+ A_2 \cos (\omega_2 t)$ with incommensurate frequencies $\omega_1$ and $\omega_2$.
Figure 4 illustrates the invisibility of the oscillating scattering potential when the modulation amplitude $R(t)$ is of the first type. The figure depicts the evolution of an initial Gaussian wave packet, propagating at an angle of 45$^o$ with respect to the primitive vectors of the Bravais lattice, which is scattered off by the Gaussian-shaped 2D potential $V_{n,m}=V_0 \exp(-n^2/w^2-m^2/w^2)$. At initial time $t=0$ the wave packet is localized at the bottom left side of the lattice, far from the scattering region, and propagates along the main lattice diagonal  toward the scattering region.  Coupled equations (26) have been numerically solved using an accurate variable-step fourth-order Runge-Kutta method on a square lattice comprising $42 \times 42$ sites; parameter values used in the simulations are $\kappa=1$, $V_0=25$, $w=2$, $\omega_1=10$, $\omega_2=2 \sqrt{18}$, and $A_1=A_2=1$. The bandwidth of the lattice is $\Delta=8 \kappa=8$. The initial condition is 
$\psi_{n,m}(0)= \mathcal{N} \exp[-(n+7)^2/9-(m+7)^2/9+i \pi (n+m)/2]$,  where $\mathcal{N}$ is the normalization constant. Panels (a), (b) and (c) of Fig.4 show the temporal evolution of the wave packet in the absence of the scattering potential [Fig.4(a)], for a modulated NH scattering potential with  $R(t)=A_1 \exp(i \omega_1 t)+A_2 \exp(i \omega_2 t)$ [Fig.4(b)], and for a modulated Hermitian scattering potential with  $R(t)=A_1 \cos(\omega_1 t)+A_2 \cos(\omega_2 t)$ [Fig.4(c)]. The full dynamical evolution is shown in the movies 1,2 and 3 of the Supplemental Material \cite{Suppl}. Clearly, in the latter case [Fig.4(c)] the potential is not invisible, and a large fraction of the incoming wave packet is scattered off by the oscillating Hermitian Gaussian-shaped potential. On the other hand, after the scattering event the wave packet in Fig.4(b) propagates as if the scattering potential were not present. This is clearly shown in Fig.4(d), which depicts the numerically-computed evolution of the error function $I(t)={\rm max}_{n,m} | \psi_{n,m}(t)-\tilde{\psi}_{n,m}|$ on a log scale, where $\psi_{n,m}(t)$ and $\tilde{\psi}_{n,m}(t)$ are the amplitudes at lattice site $(n,m)$ and at time $t$ with or without the scattering potential, respectively. Clearly, $I(t) \rightarrow 0$ as $t \rightarrow \infty$ is the signature of potential invisibility.

  \section{Conclusions and discussion}
  Wave scattering from complex potentials in non-Hermitian systems  has received a great and increasing interest in the past recent years, with the ability of tailoring the scattering properties of complex media in unprecedented ways and with the discovery of new classes of scatteringless and invisible potentials. Wave scattering is deeply influenced not only by the presence of NH potentials with gain and loss regions, that serves as source and sinks of waves, but also by the continuous or discrete spatial translational invariance of the system.\\
   Most of methods so far suggested to realize invisible or transparent potentials, both in continuous and discrete NH systems, rely on special tailoring of the potential shape, using for example the methods of supersymmetry, the spatial Kramers-Kronig relations or other related techniques. In this work we suggested a new route toward the realization of invisible potentials in NH systems with discrete spatial translational invariance, which does not require any special tailoring of the potential shape.  The key
characteristic of our new method is modulating in time {\em any arbitrary} potential shape by a complex modulation amplitude satisfying a minimal requirement, that makes any potential shape invisible. The main physical idea is that any scattering potential or defective region on a lattice, rapidly oscillating in time with only positive (or negative) frequency components, cannot scatter any propagative incoming wave into another propagative (reflected or transmitted) wave in the lattice: owing to the finite band of allowed energies in the lattice, any scattered wave is evanescent, regardless of the potential shape. As a result, any potential shape can be made invisible by making it oscillating in time under a minimal constraint.  Our approach to realize invisibile potentials in discrete systems could open up a whole new avenue for the design of synthetic media with novel scattering properties that do not rely on special engineering of material parameters. As a possible physical platform to experimentally demonstrate the new strategic method of invisibility, we suggested wave scattering in synthetic lattices based on photonic quantum walks in coupled-fiber loops, which can nowadays be routinely realized in a photonic laboratory

\appendix
\begin{widetext}
\section{Some properties of modified Laplace-Fourier integral}
{\it 1. Definition and inverse relation.} Let $f(t)$ a complex and regular function of time $t$, defined for $t \geq -t_0$ and bounded (or increasing but less than exponential) as $t \rightarrow \infty$. Indicating by $\epsilon>0$ a small positive number and $t_1$ a large positive  time instant, we define the modified Fourier-Laplace spectrum of $f(t)$, denoted by $\hat{f}^{(\epsilon)}(\omega)$, as
\begin{equation}
\hat{f}^{(\epsilon)}(\omega)=\int_{-t_0}^{t_1} dt f(t) \exp(i \omega t- \epsilon t),
\end{equation}
where $-\infty < \omega < \infty$ is the frequency. The use of the modified Fourier-Laplace transform avoids the singularities that might arise in usual Fourier analysis when $f(t)$ is bounded but non-vanishing or even weakly (non-exponentially) growing  for $t \rightarrow \infty$. 
In this work we are mainly interested in the triple limits $t_0, t_1 \rightarrow \infty$ and $\epsilon \rightarrow 0^+$, with 
\[ t_0 \epsilon \rightarrow 0 \; , \; \;  t_1 \epsilon \rightarrow \infty.  \]
This limit is justified by the need to make vanishing the boundary value terms, at $t=-t_0$ and $t=t_1$, of the functions $\phi_l(t) \exp( \epsilon t)$ when deriving the dynamical equations (10) in Fourier space, given in the main text.\\
After integration Eq.(A1) by parts, it readily follows that $\hat{f}^{(\epsilon,)}(\omega)$ decays at least as $\exp( \pm i \omega t_{1,0}) /  \omega$ for $ \omega \rightarrow \pm  \infty$, and it is thus integrable. Equation (A1) can be reversed as follows. Let us multiply both sides of Eq.(A1) by $\exp(-i \omega \tau)$ and integrate with respect to $\omega$ from $-\infty$ to $\infty$. One obtains
 \begin{equation}
\int_{-\infty}^{\infty} d \omega  \hat{f}^{(\epsilon)}(\omega) \exp (-i \omega \tau) = \int_{-\infty}^{\infty} d \omega \int_{-t_0}^{t_1} dt f(t) \exp(i \omega t- \epsilon t+i \omega \tau) 
\end{equation}
 Interchanging the integration order on the right hand side of Eq.(A4) and taking into account that
 \[
 \int_{-\infty}^{-\infty} d \omega \exp[i \omega (\tau-t)]= 2 \pi \delta(\tau-t)
 \]
one obtains
 \begin{equation}
\int_{-\infty}^{\infty} d \omega  \hat{f}^{(\epsilon)}(\omega) \exp (-i \omega \tau) = 2 \pi  \int_{-t_0}^{t_1} dt f(t) \exp(- \epsilon t)  \delta(t-\tau) 
\end{equation}
Therefore, for $-t_0 < \tau < t_1$, one has 
\begin{equation}
f(\tau)= \frac{1}{2 \pi} \exp( \epsilon \tau) \int_{- \infty}^{ \infty}  d \omega  \hat{f}^{(\epsilon)}(\omega) \exp (-i \omega \tau)
\end{equation}
which provides the inverse spectral relation.\\
\\
{\it 2. Relation to the ordinary Fourier spectrum.} When $f(t)$ admits of a Fourier spectrum $\hat{f}(\omega)$, defined in the usual way as
\begin{equation}
\hat{f}(\omega)=\int_{-\infty}^{\infty} dt f(t) \exp(i \omega t),
\end{equation}
the modified Fourier-Laplace spectrum $\hat{f}^{(\epsilon)}(\omega)$ converges to  $\hat{f}(\omega)$ in the triple limit mentioned above. In fact, we can write $\hat{f}^{(\epsilon)}(\omega)$ as the ordinary Fourier spectrum of the product between $f(t)$ and the function $h(t)$ defined by
\begin{equation}
h(t)= \left\{
\begin{array}{cc}
0 & t<-t_0 \; , \; \; t>t_1 \\
\exp( - \epsilon t) & -t_0 < t < t_1,
\end{array}
\right.
\end{equation}
i.e. 
\begin{equation}
\hat{f}^{(\epsilon)}(\omega)= \int_{-\infty}^{\infty} dt f(t) h(t) \exp(i \omega t).
\end{equation}
Using the convolution theorem of Fourier integral, one has
\begin{equation}
\hat{f}^{(\epsilon)}(\omega)= \int_{-\infty}^{\infty} d \Omega \hat{f}(\omega-\Omega) G(\Omega)
\end{equation}
where
\begin{equation}
G(\omega)=  \frac{1}{2 \pi} \hat{h}(\omega)=  \frac{1}{2 \pi}\int_{-\infty}^{\infty} dt h(t) \exp(i \omega t)=\frac{\exp[i(\omega+i \epsilon) t_1]-\exp[-i(\omega+i \epsilon) t_0] }{2 \pi i (\omega+i \epsilon)}.
\end{equation}
In the triple limit $t_0,t_1 \rightarrow \infty$, $\epsilon \rightarrow 0^+$ with $\epsilon t_0 \rightarrow 0$ and $ \epsilon t_1 \rightarrow \infty$, one has
\begin{equation}
G(\omega) \simeq - \frac{\exp(-i \omega t_0)}{2 \pi i (\omega+i \epsilon) }.
\end{equation}
Note that $|G(\omega)|=1/[2 \pi  \sqrt{\omega^2+\epsilon^2}]$ is a narrow and peaked function at around $\omega=0$, with $G(0)=1/(2 \pi \epsilon)$ diverging as $\epsilon \rightarrow 0$ and with a full-width $ \sim 2 \epsilon$ vanishing as $\epsilon \rightarrow 0$. Moreover, $G(\omega)$ is a rapidly oscillating function of $\omega$ with local zero mean for $|\omega| \gg 1/ t_0$, so that the main contribution to the integral in Eq.(A8) is obtained for $\Omega \simeq 0$. In the $\epsilon  \rightarrow 0^+$ limit, we can thus set $G(\omega)= \mathcal{A} \delta(\omega)$, where $\mathcal{A}$ is the area subtended by the function $G(\omega)$, i.e. 
\begin{equation}
\mathcal{A}=\int_{-\infty}^{\infty} d \omega G(\omega)=- \int_{-\infty}^{\infty} d \omega \frac{\exp(-i \omega t_0)}{2 \pi i (\omega+i \epsilon) }.
\end{equation}
The integral on the right hand side of Eq.(A11) can be computed in complex $\omega$ plane using the residue theorem, after closing the integration path by a semi-circumference of large radius in the lower ${\rm Im}(\omega)<0$ half plane. This yields
\begin{equation}
\mathcal{A}=\exp(-\epsilon t_0)=1
\end{equation}
where we used the limit $\epsilon t_0 \rightarrow 0$. Therefore, in the triple limit $t_0,t_1 \rightarrow \infty$ with $\epsilon t_0 \rightarrow 0$, $\epsilon t_1 \rightarrow \infty$, one has $G(\omega) \simeq \delta(\omega)$ and thus, from Eq.(A8), $\hat{f}^{(\epsilon)}(\omega) \simeq \hat{f}(\omega)$.\\
{\it 3. Convolution relation.}
Finally, let us calculate the modified Fourier-Laplace transform of the product $f(t)g(t)$, i.e. the integral
\begin{equation}
I(\omega)=\int_{-t_0}^{t_1} dt f(t) g(t) \exp(i \omega t- \epsilon t)
\end{equation}
assuming that $g(t)$ can be Fourier transformed in the usual way.
To this aim, let us write $f(t)$ and $g(t)$ in terms of their modified Fourier-Laplace and Fourier spectra, respectively, i.e. let us insert the inverse relations
\begin{equation}
f(t)= \frac{1}{2 \pi} \exp(\epsilon t )  \int_{- \infty}^{ \infty}  d \omega_1  \hat{f}^{(\epsilon)}(\omega_1) \exp (-i \omega_1 t) \; , \; \;  g(t)= \frac{1}{2 \pi} \int_{- \infty}^{ \infty}  d \omega_2  \hat{g} (\omega_2) \exp (-i \omega_2 t) 
\end{equation}
into Eq.(A13). One obtains
\begin{equation}
I(\omega)=\frac{1}{(2 \pi)^2} \int_{-t_0}^{t_1} dt  \int_{-\infty}^{\infty} d \omega_1 \int_{-\infty}^{\infty} d \omega_2  \hat{f}^{(\epsilon)}(\omega_1) \hat{g}(\omega_2) \exp[i(\omega-\omega_1-\omega_2) t].
\end{equation}
After introduction of the function
\begin{equation}
\Theta(\Omega) \equiv \int_{-t_0}^{t_1} dt \exp (i \Omega t)= \frac{\exp(i \Omega t_1)-\exp(-i \Omega t_0)}{i \Omega}
\end{equation}
one obtains
\begin{equation}
I(\omega)=\frac{1}{4 \pi^2} \int_{-\infty}^{\infty} d \omega_1 \int_{-\infty}^{\infty} d \omega_2  \hat{f}^{(\epsilon)}(\omega_1) \hat{g}(\omega_2)  \Theta (\omega-\omega_1-\omega_2).
\end{equation}
Note that in the limits $t_0, t_1 \rightarrow \infty$, $\Theta(\Omega=0)=(t_1+t_0)$ diverges, whereas for $\Omega \neq 0$ the ratio $\Theta ( \Omega) / \Theta(0)$ vanishes. Moreover, $\Theta(\Omega)$ is a rapidly oscillating function of $\Omega$ with local zero mean for $|\Omega| \gg 1/ t_0$. Therefore, in the limits $t_0, t_1 \rightarrow \infty$ one can assume $\Theta(\Omega) \simeq  \mathcal{A} \delta ( \Omega)$ in the integral on the right hand side of Eq.(A17), where $\mathcal{A}$ is the area subtended by $\Theta(\Omega)$ and given by
\begin{equation}
\mathcal{A}=\int_{-\infty}^{\infty} d \Omega \Theta (\Omega)=. \int_{-\infty}^{\infty} d \Omega \frac{\exp(i \Omega t_1)-\exp(-i \Omega t_0)}{i \Omega}.
\end{equation}
The integral on the right hand side in Eq.(A18) can be readily computed in complex $\Omega$ plane using the residue theorem, yielding $\mathcal{A}= 2 \pi$ independent of $t_0$ and $t_1$.
After letting $\Theta(\Omega) \simeq 2 \pi \delta (\Omega)$ in Eq.(A17), one finally obtains 
\begin{equation}
I(\omega) \simeq \frac{1}{2 \pi} \int_{-\infty}^{\infty} d \omega_2  \hat{f}^{(\epsilon)}(\omega-\omega_2) \hat{g}(\omega_2)
\end{equation}
which is analogous to the convolution relation of ordinary Fourier integral.

\end{widetext}

\end{document}